\documentstyle[prd,aps]{revtex}
\input epsf
\epsfverbosetrue

 \mathchardef\ScriptA="7241
 \mathchardef\ScriptB="7242 
 \mathchardef\ScriptC="7243
 \mathchardef\ScriptD="7244
 \mathchardef\ScriptE="7245
 \mathchardef\ScriptF="7246
 \mathchardef\ScriptG="7247
 \mathchardef\ScriptH="7248
 \mathchardef\ScriptI="7249
 \mathchardef\ScriptJ="724A
 \mathchardef\ScriptK="724B
 \mathchardef\ScriptL="724C
 \mathchardef\ScriptM="724D
 \mathchardef\ScriptN="724E
 \mathchardef\ScriptO="724F
 \mathchardef\ScriptP="7250
 \mathchardef\ScriptQ="7251
 \mathchardef\ScriptR="7252
 \mathchardef\ScriptS="7253
 \mathchardef\ScriptT="7254
 \mathchardef\ScriptU="7255
 \mathchardef\ScriptV="7256
 \mathchardef\ScriptW="7257
 \mathchardef\ScriptX="7258
 \mathchardef\ScriptY="7259
 \mathchardef\ScriptZ="725A

\def\mapgeq{\mathbin{\lower.3ex\hbox{$\buildrel>\over{\smash{\scriptstyle\sim}\vphantom{_x}}$}}}
\def\mapleq{\mathbin{\lower.3ex\hbox{$\buildrel<\over{\smash{\scriptstyle\sim}\vphantom{_x}}$}}}
\def\mapgeqeq{\mathbin{\lower.3ex\hbox{$\buildrel>\over{\smash{\scriptstyle\approx}\vphantom{_2}}$}}}
\def\mapleqeq{\mathbin{\lower.3ex\hbox{$\buildrel<\over{\smash{\scriptstyle\approx}\vphantom{_2}}$}}}

 \mathchardef\#="0023
 \mathchardef\$="0024
 \mathchardef\%="0025
 \mathchardef\ddash="705C
 
 \mathchardef\lwavy="336E
 \mathchardef\rwavy="336F
 \mathchardef\biglwavy="331A
 \mathchardef\bigrwavy="331B
 \mathchardef\bigglwavy="3328
 \mathchardef\biggrwavy="3329
 \mathchardef\littlesum="0350

\tighten
\draft
\begin{document}
\bibliographystyle{prsty}

\title{
Two-loop Radiative Neutrino Mechanism\\
in an $SU(3)_L\times U(1)_N$ Gauge Model
\footnote{to be published in Phys. Rev. D (2001). In {\it v2}, title \& introduction are modified and sections \& appendix are added; In {\it v3}, undefined latex citation indices in {\it v2} are removed.}
}

\author{
Teruyuki Kitabayashi$^a$
\footnote{E-mail:teruyuki@post.kek.jp}
and Masaki Yasu${\grave {\rm e}}^b$
\footnote{E-mail:yasue@keyaki.cc.u-tokai.ac.jp}
}

\address{\vspace{5mm}$^a$
{\sl Accelerator Engineering Center} \\
{\sl Mitsubishi Electric System \& Service Engineering Co.Ltd.} \\
{\sl 2-8-8 Umezono, Tsukuba, Ibaraki 305-0045, Japan}
}
\address{\vspace{2mm}$^b$
{\sl Department of Natural Science\\School of Marine
Science and Technology, Tokai University}\\
{\sl 3-20-1 Orido, Shimizu, Shizuoka 424-8610, Japan\\and\\}
{\sl Department of Physics, Tokai University} \\
{\sl 1117 KitaKaname, Hiratsuka, Kanagawa 259-1292, Japan}}
\date{TOKAI-HEP/TH-0002, June, 2000}
\maketitle

\begin{abstract}
By using the $L_e$ $-$ $L_\mu$ $-$ $L_\tau$ symmetry, we construct an $SU(3)_L\times U(1)_N$ gauge model that provides two-loop radiative neutrino masses as well as one-loop radiative neutrino masses. The generic smallness of two-loop neutrino masses leading to $\Delta m^2_\odot$ compared with one-loop neutrino masses leading to $\Delta m^2_{atm}$ successfully explains $\Delta m^2_{atm}$ $>>$ $\Delta m^2_{\odot}$ by invoking the $L_e$ $-$ $L_\mu$ $-$ $L_\tau$ breaking. The Higgs scalar ($h^+$) that initiates radiative mechanisms is unified into a Higgs triplet together with the standard Higgs scalar ($\phi^+$, $\phi^0$) to form ($\phi^+$, $\phi^0$, $h^+$), which calls for three families of lepton triplets: ($\nu^i_L$, $\ell^i_L$, $\omega^i_L$) ($i$ = 1,2,3), where $\omega^i$ denote heavy neutral leptons. The two-loop radiative mechanism is found possible by introducing a singly charged scalar, which couples to $\ell^i_R\omega^j_R$ ($i,j$ = 2,3).
\end{abstract}
\pacs{PACS: 12.60.-i, 13.15.+g, 14.60.Pq, 14.60.St\\Keywords: neutrino mass, radiative mechanism, lepton triplet}
\vspace{2mm}
\section{Introduction}
Atmospheric neutrino oscillations have been experimentally confirmed by the SuperKamiokande collaboration \cite{SuperKamiokande,RecentSK} and solar neutrinos have also been considered to be oscillating \cite{SolarNeutrino}.  These oscillation phenomena indicate new phenomenology due to 
massive neutrinos \cite{EarlyMassive} that requires new interactions beyond the conventional interactions in the 
standard model. Neutrinos can 
acquire masses through either seesaw mechanism \cite{SeeSaw} or radiative mechanism \cite{Zee,Babu,Radiative}.  
Among others, the radiative mechanism of the Zee type \cite{Zee} has recently received 
much attention \cite{ZeeType}.  The resulting structure of neutrino mass matrix on the basis of three flavors is 
characterized by 
three off-diagonal elements.  It is, then, found that in order to account for atmospheric neutrino oscillations 
by the radiative mechanism, neutrinos will exhibit bimaximal mixing \cite{BiMaxmal,NearlyBiMaximal}.  
This feature can be rephrased that there is a 
conservation of $L_e$ $-$ $L_\mu$ $-$ $L_\tau$ \cite{LeLmuLtau}, where $L_e$ ($L_\mu$ or $L_\tau$) stands 
for the electron (muon or tau) number.  

The radiative mechanism of the Zee type utilizes
\begin{enumerate}
  \item a lepton-number-violating charged lepton ($\ell_L$)-neutrino ($\nu_L$) interaction,
  \item an $SU(2)_L$-singlet charged Higgs scalar ($h^+$), and
  \item a second $SU(2)_L$-doublet Higgs scalar ($\phi'$),
\end{enumerate}
which participate in one-loop diagrams to generate Majorana neutrino masses.  
The simplest extension of the standard gauge group, $SU(2)_L \times U(1)_Y$, so as to include these extra 
ingredients, especially for $h^+$, is to employ an $SU(3)_L\times U(1)_N$ gauge group \cite{SU3U1}.  
The charged Higgs 
scalar, $h^+$, can be identified with the third member of an $SU(3)_L$-triplet Higgs scalar and is unified in to 
($\phi^+$, $\phi^0$, $h^+$) together with the standard Higgs doublet ($\phi^+$, $\phi^0$) \cite{OkamotoYasue}. 
It is also known that another advantage to use $SU(3)_L\times U(1)_N$ 
lies in the fact that three families of quarks and leptons are predicted if the anomaly free condition on 
$SU(3)_L\times U(1)_N$ and the asymptotic free condition on $SU(3)_c$ are imposed.  
These plausible properties have pushed us to examine how to implement one-loop radiative mechanism within 
the $SU(3)_L\times U(1)_N$ framework \cite{OkamotoYasue,EarlyRadiativeNu,RadiativeNu,TreeNu}.

In this report, we further study radiative 
mechanism based on two-loop diagrams \cite{BabuType} in an $SU(3)_L\times U(1)_N$ model with ($\phi^+$, $\phi^0$, $h^+$) as a Higgs triplet \cite{OtherVariant}, which in turn requires three 
families of heavy neutral leptons to be denoted by $\omega^i$ ($i$ = 1,2,3) \cite{HeavyLepton}, which are contained in 
lepton triplets of ($\nu^i_L$, $\ell^i_L$, $\omega^i_L$). 
It is anticipated that the interactions described by one- 
and two-loop diagrams can be used to generate the observed atmospheric and 
solar neutrino oscillations characterized by $\Delta m_{atm}^2>>\Delta m_\odot^2$ 
as have been stressed in Ref.\cite{OneTwoLoop}. 
Namely, one-loop radiative mechanism controls atmospheric neutrino oscillations while further suppressed 
effects from two-loop radiative mechanism provide solar neutrino oscillations.
To activate two-loop radiative mechanism, an extra scalar called $k^+$ is introduced, which 
will couple to a lepton-number-violating $\ell_R$-$\omega_R$ \cite{Simplest}
and $\rho^{\prime +\dagger}{\bar \eta}^0k^+$, where $\rho^{\prime +}$ and ${\bar \eta}^0$ are 
to be introduced in Eq.(\ref{Eq:HiggsEtaRhoChi}). 

\section{Model}   
The present $SU(3)_L\times U(1)_N$ gauge model is specified by the $U(1)_N$ quantum number, $N/2$, which 
is related to the hypercharge, $Y$, as $Y$ = $-\lambda^8/\sqrt{3}$ + $N$. The electric charge, 
$Q_{em}$, is, thus, given by $Q_{em}$ = ($\lambda^3$ + $Y$)/2, where $\lambda^a$ is the $SU(3)$ generator 
with Tr($\lambda^a\lambda^b$) = 2$\delta^{ab}$ ($a,b$ = 1$\sim$8).
The particle content can be summarized as follows:
\begin{eqnarray}\label{Eq:LeptonTriplets}
& \psi^i_L = \left( \nu^i, \ell^i, \omega^i\right)^T_L:({\bf 3}, -\frac{1}{3}), \ \ \ 
\ell^i_R:~ ({\bf 1}, -1), \ \ \ \omega^i_R:~ ({\bf 1}, 0),
\end{eqnarray}
for leptons, where $\omega^i$ stand for three neutral heavy leptons,
\begin{eqnarray}
&& Q^1_L  = \left( u^1,  d^1,  u^{\prime 1}\right)^T_L:({\bf 3}, \frac{1}{3}), \ \ \  
Q^{i=2,3}_L  =  \left( d^i,  -u^i,  d^{\prime i}\right)^T_L:({\bf 3^\ast}, 0), 
\nonumber \\
&& u^{1,2,3}_R:({\bf 1}, \frac{2}{3}), \ \ \ d^{1,2,3}_R:({\bf 1}, -\frac{1}{3}), \ \ \  
u^{\prime 1}_R:({\bf 1}, \frac{2}{3}), \ \ \ d^{\prime 2,3}_R:  ({\bf 1}, -\frac{1}{3}).  
\label{Eq:ExtraQuarkTriplets}
\end{eqnarray}
for quarks.  The values in the parentheses denote quantum numbers 
based on the ($SU(3)_L$, $U(1)_N$) - symmetry.  It is obvious that the pure 
$SU(3)_L$ anomaly vanishes since there is an equal number of  triplets of quarks and leptons and 
antitriplets of quarks.  Other anomalies are also cancelled.  
Our Higgs scalars are given by
\begin{enumerate}
	\item three triplets ($\eta$, $\rho$, $\chi$) that provides three members of the quark triplets as well as ordinary and heavy leptons, and one duplicate of $\rho$ ($\rho^\prime$) that initiates one-loop radiative mechanism as advocated by Ref.\cite{Duplicate}, 
\begin{eqnarray}
&& \rho = (\rho^+, \rho^0, {\bar \rho^+})^T:({\bf 3}, \frac{2}{3}),  \ \ \
\rho' = (\rho'^+, \rho'^0, {\bar \rho'^+})^T:({\bf 3}, \frac{2}{3}),  
\nonumber \\
&& \eta = (\eta^0, \eta^-, {\bar \eta}^0)^T:({\bf 3}, -\frac{1}{3}),  \ \ \
\chi = ({\bar \chi}^0, \chi^-, \chi^0)^T:({\bf 3}, -\frac{1}{3}), 
\label{Eq:HiggsEtaRhoChi}
\end{eqnarray}
from which quarks and leptons acquire masses through their vacuum expectation values (VEV's):
\begin{equation}\label{Eq:VEVs}
\langle 0|\eta |0\rangle = ( v_\eta, 0, 0)^T, \ \    
\langle 0|\rho |0\rangle = (0,  v_\rho, 0)^T, \ \   
\langle 0|\chi |0\rangle = (0,  0, v_\chi)^T,
\end{equation}
where the orthogonal choice of these VEV's and the absence of $\langle 0|\rho' |0\rangle$ will be 
guaranteed by appropriate Higgs interactions to be introduced;
	\item one singly charged scalar that initiates two-loop radiative mechanism.
\begin{equation}\label{Eq:Singlet}
k^+: ({\bf 1}, 1),
\end{equation}
which will couple to $\ell^i_R\omega^j_R$ ($i,j$ = 2,3).
\end{enumerate}

The model becomes acceptable for the present discussions if it satisfies the following requirements that
\begin{itemize}
	\item the Higgs scalar, $\rho$, acquire a VEV and be responsible for creating mass terms for quarks and leptons,
	\item the Higgs scalar, $\rho'$, acquire no VEV and be responsible for lepton-number-violating interactions,
	\item flavor-changing interactions due to Yukawa interactions well be suppressed.
\end{itemize}
Since lepton-number-violating interactions can be described by 
$\epsilon^{\alpha\beta\gamma}\psi^i_{L\alpha}\psi^j_{L\beta}\rho_\gamma$, which will generate 
the $\nu^i_L$-$\omega^j_L$ mixing owing to $\langle 0|\rho_2 |0\rangle$ $\neq$ 0, leading to a tree-level 
neutrino mass matrix \cite{Duplicate}.  Therefore, this coupling of $\rho$ should be avoided 
in the radiative mechanism.  
The model contains quarks with the same charge, whose mass terms can be generated by $\eta$ and $\chi$ between 
$Q^1_L$ and up-type quarks and by $\eta^\dagger$ and $\chi^\dagger$ between $Q^{2,3}_L$ and down-type quarks.  
To achieve flavor-changing-neutral-currents (FCNC)
suppression, Yukawa interactions must be constrained such that 
a quark flavor gains a mass from only one Higgs field \cite{FCNC,FCNCSU3}. 

All these restrictions can be realized by requiring interactions be invariant under a discrete transformation 
based on $Z_4$, which is given by
\begin{eqnarray}\label{Eq:LeptonDiscrete}
& &  \psi^{1,2,3}_L \rightarrow i\psi^{1,2,3}_L, \ \ \ 
\ell^{1,2,3}_R \rightarrow \ell^{1,2,3}_R, \ \ \ 
\omega^{1,2,3}_R \rightarrow -i\omega^{1,2,3}_R,
\end{eqnarray}
for leptons, 
\begin{eqnarray}\label{Eq:QuarkDiscrete}
& &  Q^1_L \rightarrow iQ^1_L, \ \ \ 
Q^{2,3}_L \rightarrow -iQ^{2,3}_L, \ \ \ 
u^{1,2,3}_R \rightarrow u^{1,2,3}_R, \ \ \ 
d^{1,2,3}_R \rightarrow d^{1,2,3}_R, \ \ \ 
\nonumber \\
& & u^{\prime 1}_R \rightarrow -iu^{\prime 1}_R, \ \ \ 
d^{\prime 2,3}_R \rightarrow id^{\prime 2,3}_R, 
\end{eqnarray}
for quarks, and
\begin{eqnarray}\label{Eq:HiggsDiscrete}
& &  \eta \rightarrow i\eta, \ \ \ 
\rho \rightarrow i\rho, \ \ \ 
\rho^\prime \rightarrow -\rho^\prime, \ \ \ 
\chi \rightarrow -\chi, \ \ \ 
k^+ \rightarrow ik^+,
\end{eqnarray}
for Higgs scalars.  In addition to the discrete symmetry, we also impose the $L_e$ $-$ $L_\mu$ $-$ $L_\tau$ 
($\equiv$ $L^\prime$) 
conservation on our interactions to reproduce the observed atmospheric neutrino oscillations. 
The quantum number, $L^\prime$, is assigned to be 
0 for ($\eta$, $\rho$, $\rho'$, $\chi$), 
1 for ($\psi^1_L$, $\ell^1_R$, $\omega^1_R$), 
$-$1 for ($\psi^{2,3}_L$, $\ell^{2,3}_R$, $\omega^{2,3}_R$) and 
2 for $k^+$ and similarly for quarks. The non-vanishing lepton number is carried by $k^+$ 
with $L$=$-$2 and as well as by leptons with $L$=1.

Yukawa interactions are controlled by the following lagrangian:
\begin{eqnarray}
-{\cal L}_Y  & = &  
\frac{1}{2}\epsilon^{\alpha\beta\gamma}\sum_{i=2,3} f_{[1i]}
{\overline {\left( \psi_{\alpha L}^1 \right)^c}}\psi^i_{\beta L}\rho'_\gamma 
+\sum_{i=1,2,3}{\overline {\psi^i_L}}\left( f^i_{\ell}\rho\ell_R^i+f^i_{\omega}\chi\omega^i_R\right)
\nonumber \\
& & +\sum_{i,j=2,3}f^{ij}_h{\overline {\left( \ell^i_R \right)^c}}\omega^j_Rk^+ 
 +  {\overline {Q^1_L}}\left(
\eta U^1_R + \rho D^1_R+\chi U^{\prime 1}_R
\right) 
\nonumber \\
& &  + \sum_{i=2,3}{\overline {Q^i_L}}\left(\rho^\ast U^i_R
+\eta^\ast D^i_R  + \chi^\ast D^{\prime i}_R
\right) 
+ {\rm (h.c.)},
\label{Eq:Yukawa}
\end{eqnarray}
where $f$'s denote the Yukawa couplings and $f_{[ij]}$ = $-f_{[ji]}$.  Right-handed quarks are 
specified by $U^i_R$ = $\sum_{j=1}^3f^i_{uj}u^j_R$, $D^i_R$ = $\sum_{j=1}^3f^i_{dj}d^j_R$, 
$U^{\prime i}_R$ = $f^i_{u^\prime 1}u^{\prime 1}_R$ and 
$D^{\prime i}_R$ = $\sum_{j=2}^3f^i_{d^\prime j}d^{\prime j}_R$.  
The charged lepton and heavy lepton mass matrices are taken to be diagonal.  
We here note that
\begin{enumerate}
	\item the $L$-breaking is supplied by ${\overline {\left( \psi_L^1 \right)^c}}\psi^i_L\rho'$ but all other interactions conserve both $L$ and $L^\prime$,
	\item the coupling of $k^+$ to the first family is forbidden by the $L^\prime$ symmetry, 
	\item the absence of $\chi^\ast D^i_R$ and $\eta^\ast D^{\prime i}_R$ for $Q^{2,3}_L$ and of $\chi U^1_R$ and $\eta U^{\prime 1}_R$ for $Q^1_L$ ensures the suppression of FCNC.
\end{enumerate}
The Higgs interactions are described by
self-Hermitian terms composed of $\phi_\alpha\phi^\dagger_\beta$ ($\phi$ = $\eta$, $\rho$, 
$\chi$, $k^+$) and by the non-self-Hermitian terms in
\begin{eqnarray}\label{Eq:Conserved}
V_0 & = &  
\lambda_0\epsilon^{\alpha\beta\gamma}\eta_\alpha\rho_\beta\chi_\gamma + 
\lambda_1\left( \rho^\dagger\eta \right) \left( \chi^\dagger\rho'\right)  + 
\lambda_2\left( \rho^\dagger\rho' \right) \left( \chi^\dagger\eta\right) + {\rm (h.c.)},
\end{eqnarray}
which conserves both $L$ and $L^\prime$, where $\lambda_{0,1,2}$ represent coupling constants.  
To account for solar neutrino oscillations, the breaking of the $L^\prime$-conservation should be included 
and is assumed to be furnished by 
\begin{eqnarray}\label{Eq:Broken}
V_b & = & \mu_b\rho^{\prime\dagger} \eta k^+ + {\rm (h.c.)},
\end{eqnarray}
which also breaks the $L$-conservation, 
where $\mu_b$ represents a breaking scale of the $L^\prime$-conservation.  
The orthogonal choice of VEV's of $\eta$, $\rho$ and $\chi$ as in 
Eq.(\ref{Eq:VEVs}) is supported by $V_0$ if $\lambda_0$ $<$ 0.  
It is because $V_0$ gets lowered if $\eta$, $\rho$ and $\chi$ develop VEV's.  So, one can choose 
VEV's such that $\langle 0|\eta_1 |0\rangle$ $\neq$ 0, $\langle 0|\rho_2 |0\rangle$ $\neq$ 0 and 
$\langle 0|\chi_3 |0\rangle$ $\neq$ 0. However, the similar coupling of $\eta\rho^\prime\chi$ is 
forbidden by the $Z_4$ symmetry; therefore, the requirement of $\langle 0|\rho^\prime |0\rangle$ = 0 
is not disturbed.  Also forbidden is the dangerous term of $\rho^\dagger\rho^\prime$ that would 
induce a non-vanishing VEV of $\rho^\prime$.  Finally, it should be noted that the $L$- and/or 
$L^\prime$-breakings are supplied by the $\psi\psi\rho^\prime$- and $\rho^\prime\eta k^+$-terms 
but all others conserve both $L$ and $L^\prime$.

\section{Radiative Neutrino Masses}
The explicit form of Yukawa interactions relevant for the radiative neutrino mechanism is given by
\begin{eqnarray}
&&\sum_{i=2,3}\left\{f_{[1i]}\left[
\left( {\overline {\ell^{c1}_R}}\nu^i_L - {\overline {\nu^{1c}_R}}\ell^i_L \right){\bar \rho}^{\prime +} 
+\left( {\overline {\omega^{c1}_R}}\ell^i_L - {\overline {\ell^{c1}_R}}\omega^i_L \right) \rho^{\prime +} 
\right.\right.
\nonumber \\
&& \left.\left.
\hspace{50pt}+\left( {\overline {\nu^{1c}_R}} \omega^i_L - {\overline {\omega^1_R}} \nu^i_L \right) 
\rho^{\prime 0} 
\right]-f^{ij}_h{\overline {\ell^{ci}_L}}k^+\omega^j_R\right\} 
-f^i_{\ell i}\left( {\overline {\nu^i_L}}\ell^i_R\rho^+ 
+ {\overline {\ell^i_L}}\ell^i_R\rho^0 \right) 
\nonumber \\
&&\hspace{80pt}- f^i_{\omega i}\left( {\overline {\nu^i_L}}\omega^i_R\chi^-
+ {\overline {\omega^i_L}}\omega^i_R\chi^0 \right) 
+ {\rm (h.c)}.
\label{Eq:NuMasses}
\end{eqnarray}
The combined use of these interactions with $V_0$ yields one-loop diagrams for $L^\prime$-conserving Majorana 
neutrino mass terms as shown in Fig.\ref{Fig_1ab} (a) and (b), which correspond to 
\begin{equation}\label{Eq:EffectiveCoupling1}
\left( \eta^\dagger\psi^1_L\right) \epsilon^{\alpha\beta\gamma}\rho_\alpha\chi_\beta\psi^{2,3}_{\gamma L}
+
\left( \eta^\dagger\psi^{2,3}_L\right) \epsilon^{\alpha\beta\gamma}\rho_\alpha\chi_\beta\psi^1_{\gamma L},
\end{equation}
leading to one-loop radiative masses to be denoted by $m^{(1)}_{12,13}$.  On the other hand,
 the $L^\prime$-violating $V_b$ provides two-loop 
diagrams as depicted in Fig.\ref{Fig_2}, which correspond to an effective coupling
\begin{equation}\label{Eq:EffectiveCoupling2}
\epsilon^{\alpha\beta\gamma}\epsilon^{\alpha'\beta'\gamma'}
\psi^1_{\alpha L}\rho_\beta\chi_\gamma
\psi^1_{\alpha' L}\rho_{\beta'}\chi_{\gamma'},
\end{equation}
leading to two-loop radiative masses to be denoted by $m^{(2)}_{11}$.

The resulting neutrino mass matrix is found to be
\begin{eqnarray}\label{Eq:NUmass}
M_\nu = \left( \begin{array}{ccc}
  m^{(2)}_{11}&  m^{(1)}_{12}&   m^{(1)}_{13}\\
  m^{(1)}_{12}&  0&  0\\
  m^{(1)}_{13}&  0&  0\\
\end{array} \right),
\end{eqnarray}
where $m^{(1,2)}_{ij}$ are calculated to be
\begin{eqnarray}
m^{(1)}_{1i}  &=&  f_{[1i]} v_\eta v_\rho v_\chi
\left[ 
 \lambda_1\frac{m^2_{\ell_i}F(m^2_{\ell_i}, m_{{\bar \rho}^{\prime 0}}^2, m_{\rho^+}^2)-
m^2_{\ell_1}F(m^2_{\ell_1}, m_{{\bar \rho}^{\prime 0}}^2, m_{\rho^+}^2)}{v^2_\rho} \right.
\nonumber \\
  &&+\left. \lambda_2\frac{m^2_{\omega_i}F(m^2_{\omega_i}, m_{\rho^{\prime 0}}^2, m_{{\bar \chi}^0}^2)
  -m^2_{\omega_1}F(m^2_{\omega_1}, m_{\rho^{\prime 0}}^2, m_{{\bar \chi}^0}^2)}{v^2_\chi} 
\right] , 
\label{Eq:MatrixEntryIJ1} \\
m^{(2)}_{11}  &=& -2\sum_{ij=2,3}\lambda_2 f_{[1i]}f_{[1j]}f^{ij}_hm_{\ell_i} m_{\omega_j}
\mu_b v_\rho v_\chi I_{two-loop}.
\label{Eq:MatrixEntryIJ2}
\end{eqnarray}
The mass parameters of $m_{\ell_i}$ ($\equiv$ $f^i_{\ell i}v_\rho$) 
and $m_{\omega_i}$ ($\equiv$ $f^i_{\omega i}v_\chi$) are, respectively, the mass of the $i$-th 
charged lepton and the mass of the $i$-th heavy neutral lepton and 
masses of Higgs scalars are denoted by the subscripts in terms of their fields.  
The function of $F$ and the two-loop integral of $I_{two-loop}$ are, respectively, given by
\begin{eqnarray}
&&F(x, y, z) = \frac{1}{16\pi^2}
\left[
\frac{x\log x}{(x-y)(x-z)}
+\frac{y\log y}{(y-x)(y-z)}
+\frac{z\log z}{(z-y)(z-x)}
\right],
\nonumber \\
\label{Eq:F_x_y}
\end{eqnarray}
and 
\begin{eqnarray}
&&I_{two-loop} = 
\frac{
G\left( {m_\ell^2 ,m_{{\bar \rho}^{\prime+}}^2 ,m_k^2 } \right)
}
{
m_k^2
}
\cdot
\frac{
G\left( {m_\omega^2 ,m_{{\bar \eta}^0}^2 ,m_k^2 } \right)
-G\left( {m_\omega^2 ,m_{{\bar \rho}^{\prime 0}}^2 ,m_k^2 } \right)
}
{
m_{{\bar \eta}^0}^2-m_{{\bar \rho}^{\prime 0}}^2
}
\label{Eq:I_two-loop}
\end{eqnarray}
with
\begin{eqnarray}\label{Eq:G_x_y}
G\left( x,y,z \right) = \frac{1}{16\pi ^2 }\frac{x\ln \left( x/z \right) - 
y\ln \left( y/ z\right)}{x-y}.
\end{eqnarray}
The explicit form of $I_{two-loop}$ of Eq.(\ref{Eq:I_two-loop}) is possible to obtain if 
the squared mass of $k^+$,  $m_k^2$, is much greater than the squared masses shown 
inside parentheses in (\ref{Eq:I_two-loop}), namely, $m_k^2$ $\gg$ 
$m^2_{\ell^{2,3}}$, $m^2_{\omega^{2,3}}$, $m^2_{\bar \rho^{\prime 0,+}}$, $m^2_{\bar \eta^0}$. 
The outline of its derivation is shown in the Appendix.

The neutrino oscillations are characterized by $\Delta m^2_{atm}$ and $\Delta m^2_{\odot}$, which are given by 
\begin{equation}\label{Eq:NUmass2}
\Delta m^2_{atm} = m^{(1)2}_{12} + m^{(1)2}_{13} (\equiv m^2_\nu), \ \ \
\Delta m^2_{\odot} = 2 m_\nu \vert m^{(2)}_{11}\vert,
\end{equation}
where the anticipated relation of $\vert m^{(2)}_{11}\vert$ $\ll$ $\vert m^{(1)}_{1i}\vert$ 
for $i$ = 2,3 has been used. 
To enhance the bimaximal structure of $M_\nu $ in Eq.(\ref{Eq:NUmass}), we assume the equality of 
$m_{\omega_2}$ = $m_{\omega_3}$ together with $f_{[12]}$ = $f_{[13]}$ in the heavy lepton sector and 
$m_{\omega_1}$ $\neq$ $m_{\omega_{2,3}}$ to yield nonvanishing radiative corrections due to the 
heavy-lepton-exchanges. 
Tiny contributions from the ordinary lepton sector yield the deviation from the bimaximal structure. 

To get order of magnitude estimates of our parameters, we first use $\Delta m^2_{atm}$ 
$\sim$ $3\times 10^{-3}$ eV$^2$ for atmospheric neutrino oscillations\cite{SuperKamiokande,RecentSK,RecentAtm} to 
mainly fix the lepton-number-violating coupling of $f_{[1i]}$ ($i$=2,3). 
It is then shown that solar neutrino oscillations are to be characterized by $\Delta m^2_{\odot}$ 
$\sim$ $2 \times 10^{-10}$ eV$^2$.  To be more specific, we use the following parameters to compute 
$\Delta m^2_{atm}$ and $\Delta m^2_{\odot}$ in Eq.(\ref{Eq:NUmass2}):
\begin{itemize}
\item $\sqrt{v^2_{\eta}+v^2_{\rho}}$ = $v_{weak}$, where $v_{weak}$ = $( 2{\sqrt 2}G_F)^{-1/2}$ = 174 GeV, 
since these are the sources of weak boson masses, 
where $v_\rho$ = $v_{weak}$ and $v_\eta$ = $v_\rho$/10 are taken,
\item $v_\chi$ = 10$v_{weak}$ so that exotic particles acquire sufficiently large masses,
\item $m_{\omega_{2,3}}$ = $ev_\chi$ with $m_{\omega_{2,3}}-m_{\omega_1}$ = $m_{\omega_{2,3}}/10$, 
where $e$ stands for the electromagnetic coupling, 
\item $m_\eta$ = $m_{\rho^\prime}$ = $v_\rho$ and 
$m_\chi$ = $m_k$ = $v_\chi$,
\item the following dimensionless couplings related to $L$- and $L^\prime$-conserving interactions are 
kept to be order unity: $\lambda_1$ = $\lambda_2$ = $f_h^{ij}$ = 1 ($i,j$ = 2,3),
\item the dimensionless couplings related to $L$- or $L^\prime$-violating interactions are 
considered to be small: $f_{[1i]}$ $\ll$ 1 and $\mu_b$ $\ll$ $v_\chi$, where $f_{[1i]}$ = 
$10^{-7}$ (to reproduce $\Delta m^2_{atm}$) and $\mu_b$ = $ev_\chi$ are taken.
\end{itemize}
The adopted magnitude of $f_{[1i]}$ (=10$^{-7}$) can be seen from the rough 
estimate of $m_{1i}^{(1)}$. By noticing that $F(m^2_\omega,m^2_{\rho^{\prime0}},m_k^2)$ 
$\sim$ $F(m^2_\omega,0,m_k^2)$ = $\ln(m_k^2/m^2_\omega)/(16\pi^2m_k^2)$,
we obtain
$m_{1i}^{(1)}$ $\sim$ $5f_{[1i]}\times 10^{-4}~{\rm GeV}$ ($i$=2,3). 
This estimate certainly gives $f_{[1i]}$ $\sim 10^{-7}$ 
in order to obtain $m_{1i}^{(1)}$ = $\sqrt{\Delta m^2_{atm}/2}$ $\sim$ $4\times 10^{-2}$ eV for 
$\Delta m_{atm}^2$ $\sim$ 3$\times 10^{-3}$ eV$^2$.  
The numerical computation by 
the exact formula of $I_{two-loop}$ involving Eq.(\ref{Eq:Result}) shown in the Appendix 
reproduces $\Delta m^2_{atm}$ = $3 \times 10^{-3}$ eV$^2$ and yields 
$\Delta m^2_{\odot}$ = $2 \times 10^{-10}$ eV$^2$. The approximate bimaximal 
mixing is characterized by $\sin2\theta$ = 0.97 reflecting the contributions from the charged lepton-exchanges 
to $m_{1i}^{(1)}$, where $\sin 2\theta$=1 corresponds to the bimaximal mixing.  
One can observe that the estimated $\Delta m^2_{\odot}$ lies in the allowed region of the 
observed $\Delta m^2_{\odot}$ for the vacuum oscillations \cite{SolarNeutrino,VOComment}.  
Another case of $k^+$ with $L^\prime$ = $-2$ that couples to $e_Re_R$ 
yields $\Delta m^2_{\odot}$ $\sim$ $2 \times 10^{-10}(m_e/m_\tau )$ eV$^2$, which is inconsistent 
with the solar neutrino oscillation data.

\section{Summary}
Summarizing our discussions, we have clarified how two-loop radiative mechanism is implemented in 
the $SU(3)_L$ $\times$ $U(1)_N$ model with the 
lepton triplets ($\nu^i$, $\ell^i$, $\omega^i$)$^T$, where Zee's scalar, $h^+$, that has 
initiated radiative mechanisms is unified into ($\phi^+$, $\phi^0$, $h^+$) together with the standard Higgs 
doublet, ($\phi^+$, $\phi^0$).
The interactions respecting the 
$L_e$ $-$ $L_\mu$ $-$ $L_\tau$ (=$L^\prime$) conservation are responsible for the one-loop radiative corrections, 
which generate atmospheric neutrino oscillations, while solar neutrino oscillations are induced, 
via two-loop radiative corrections, by 
the explicit $L^\prime$-breaking, which is supplied by Higgs interactions involving a charged scalar 
($k^+$) with $L^\prime$ = 2, 
which couples to $\tau^-_R\omega^{2,3}_R$.  One-loop Majorana masses receive dominant contributions from the 
heavy-lepton-exchanges 
and the bimaximal structure is preferred by the approximate degeneracy between 
$\omega^2$- and $\omega^3$-masses.  It is, then, shown that, for $\Delta m^2_{atm}$ $\sim$ 
$3\times 10^{-3}$ eV$^2$, $\Delta m^2_\odot$ of order $10^{-10}$ eV$^2$ is obtained as the vacuum solution. 
The approximate bimaximal structure is characterized by $\sin2\theta$ = 0.97.

\begin{center}
{\bf ACKNOWLEDGMENTS}
\end{center}

One of the authors (M.Y.) is grateful to M. Matsuda for comments on the usefulness of two-loop diagrams.  
The work of M.Y. is supported by the Grant-in-Aid for Scientific Research No 12047223 from the 
Ministry of Education, Science, Sports and Culture, Japan.

\noindent
{\bf \center Appendix\\}
In this Appendix, we describe the detailed evaluation of the two-loop integral of Eq.(\ref{Eq:I_two-loop}).
The relevant integration corresponding to Fig.\ref{Fig_2} reads 
\begin{eqnarray}\label{Eq:Start}
I_{two-loop} &=& \int \!\!
\frac{d^4 k}{\left( 2\pi  \right)^{4}}
\frac{d^4 q}{\left( 2\pi \right)^{4}}
\frac{1}{k^2  - m_\ell^2 }
\frac{1}{k^2  - m_{\rho^\prime}^2 }
\frac{1}{q^2  - m_\omega^2 }
\nonumber \\
&&\quad\quad\quad\quad\quad\quad
\cdot\frac{1}{q^2  - m_\eta^2 }
\frac{1}{q^2  - m_{\rho^\prime}^2 }
\frac{1}{\left( {k - q} \right)^2  - m_k^2 }.
\end{eqnarray}
By using that
\begin{eqnarray}\label{Eq:Separate}
&& \frac{1}{q^2 - m_\eta^2}\frac{1}{q^2  - m_{\rho^\prime}^2}=
\frac{1}{m_\eta^2-m_{\rho^\prime}^2}\left( \frac{1}{q^2  - m_\eta^2}-\frac{1}{q^2  - m_{\rho^\prime}^2}\right),
\end{eqnarray}
we find that
\begin{eqnarray}\label{Eq:TwoLoop0}
&&I_{two-loop} = \frac{J\left(m_\eta^2 \right)-J\left(m_{\rho^\prime}^2 \right)}{m_\eta^2-m_{\rho^\prime}^2},
\end{eqnarray}
where
\begin{eqnarray}\label{Eq:JStart}
J\left(m^2 \right) &=& \int \!\!
\frac{d^4 k}{\left( 2\pi  \right)^{4}}
\frac{d^4 q}{\left( 2\pi \right)^{4}}
\frac{1}{k^2  - m_\ell^2 }
\frac{1}{k^2  - m_{\rho^\prime}^2 }
\frac{1}{q^2  - m_\omega^2 }
\frac{1}{q^2  - m^2 }
\frac{1}{\left( {k - q} \right)^2  - m_k^2 }.
\nonumber \\
\end{eqnarray}
After the integration over $k$ is performed, we reach
\begin{eqnarray}\label{Eq:Result}
&&J\left(m^2 \right) = \int^1_0 \!\! dx\int^1_0 \!\! ydy
\frac{i}{16\pi ^2 \left[ y\left( {1 - y} \right) \right]}
I_{one-loop}\left( m_\omega^2,m^2,M(x,y)^2\right)
\end{eqnarray}
with 
\begin{equation}\label{Eq:xyMass}
M(x,y)^2  = \frac{m_k^2  - \left( m_k^2  - m_{\rho^\prime}^2 \right)y - \left( m_{\rho^\prime}^2  - m_\ell^2 \right)xy}{y\left( 1 - y \right)},
\end{equation}
where $I_{one-loop}$ represents the one-loop integral expressed by $F$ of Eq.(\ref{Eq:F_x_y}):
\begin{eqnarray}\label{Eq:OneLoopIntegral}
I_{one-loop}(a,b,c)&=& \int \!\! \frac{d^4 q}{\left( {2\pi } \right)^{4} }
\frac{1}{q^2  - a}\frac{1}{q^2  - b}\frac{1}{q^2  - c} = -iF\left( a,b,c\right).
\end{eqnarray}

Under the approximation of $m_k^2$ $\gg$ $m^2,m_{\ell, \omega, \rho^\prime}^2$, 
following the estimation of $J(m^2)$ shown in the Appendix of Ref.\cite{Appendix}, we find that
\begin{eqnarray}\label{Eq:FinalResult}
J\left(m^2 \right) = \frac{{G\left( {m_\ell^2 ,m_{\rho^\prime}^2 ,m_k^2 } \right)
G\left( {m_\omega^2 ,m^2 ,m_k^2 } \right)}}{{m_k^2 }},
\end{eqnarray}
where
\begin{eqnarray}\label{Eq:FinalG_x_y}
G\left( x,y,z \right) = \frac{1}{16\pi ^2 }\frac{x\ln \left( x/z \right) - 
y\ln \left( y/ z\right)}{x-y}.
\end{eqnarray}
Collecting these expressions, we finally obtain Eq.(\ref{Eq:I_two-loop}).

\noindent
{\bf \center Figure Captions\\}
\begin{figure}
\caption{One loop radiative diagrams for $\nu^1$-$\nu^i$ ($i$=2,3) via (a) charged leptons and 
(b) heavy leptons.}
\label{Fig_1ab}
\end{figure}

\begin{figure}
\caption{Two loop radiative diagrams for $\nu^1$-$\nu^1$.}
\label{Fig_2}
\end{figure}

\newpage
\centerline{\epsfbox{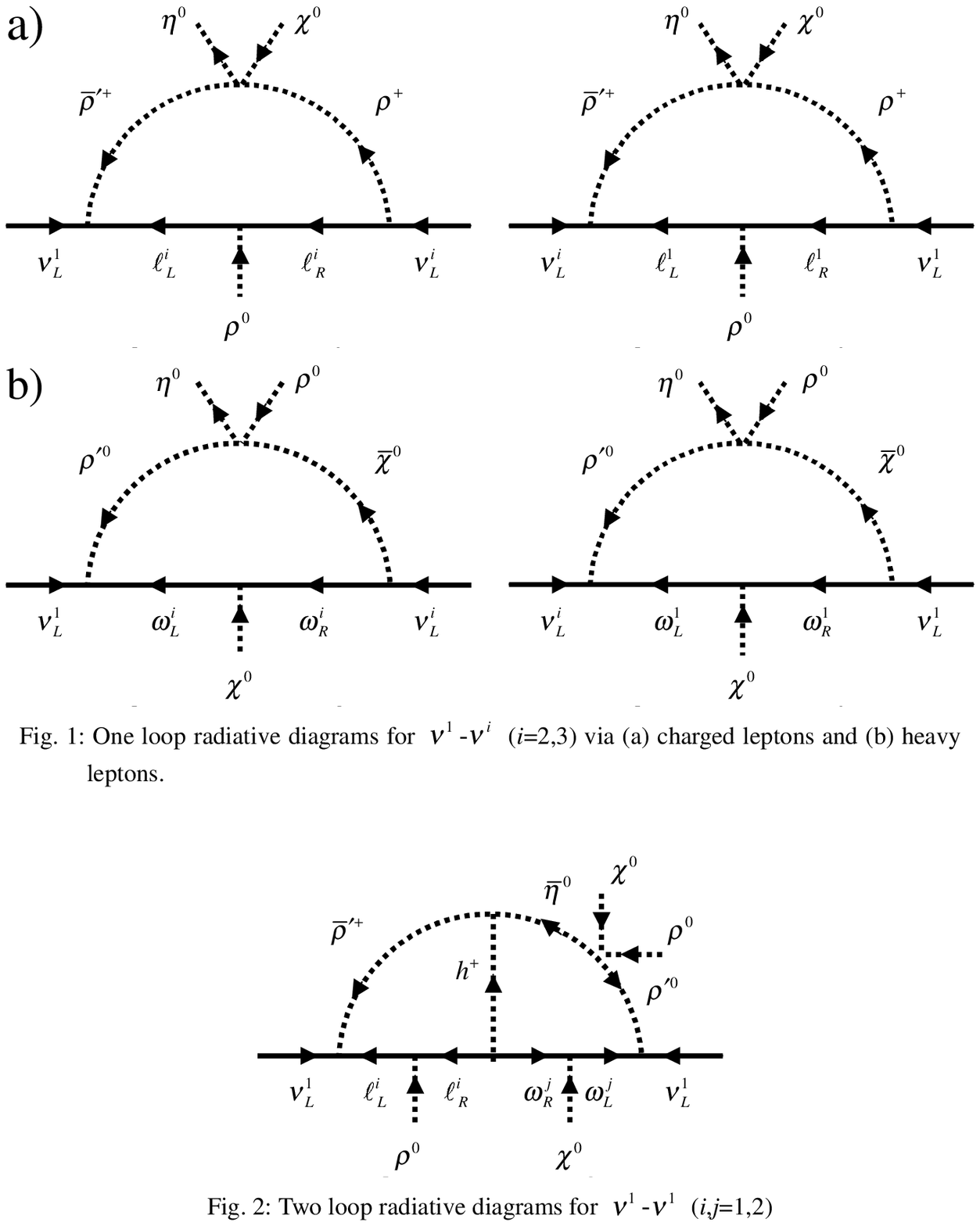}}

\begin{references}
\bibitem{SuperKamiokande} SuperKamiokande collaboration, Y. Fukuda {\it et al.}, 
Phys.  Rev. Lett. {\bf 81}, 1562 (1998); Phys. Lett. B {\bf 433}, 9 (1998) and {\bf 436}, 33 (1998). 
See also K. Scholberg, hep-ex/9905016 (May, 1999).

%
\bibitem{RecentSK}
    Y. Takeuchi, Talk given at 
    \textit{the 30th Int. Conf. on High Energy Physics (ICHEP2000)}, 
    27 Jul. - 2 Aug., Osaka, Japan. 

\bibitem{SolarNeutrino}
See for exampe, J. N. Bahcall, P. I. Krastev and A.Yu.Smirnov, 
    Phys. Rev. D \textbf{58}, 096016 (1998); \textit{ibid}. \textbf{60}, 093001 (1999);
    J. N. Bahcall, hep-ph/0002018 (Feb, 2000);
    M. C. Gonzalez-Garcia, P. C. de Holanda, C. Pena-Garay and J.W.F. Valle,
    Nucl. Phys. B \textbf{573}, 3 (2000).  

%
\bibitem{EarlyMassive} Z. Maki, M. Nakagawa and S. Sakata, Prog. Theor. Phys. {\bf 28}, 870 (1962). 
See also  B. Pontecorvo, JETP (USSR) {\bf 34}, 247 (1958);  
B. Pontecorvo, Zh. Eksp. Teor. Piz. {\bf 53}, 1717 (1967);
V. Gribov and B. Pontecorvo, Phys. Lett. {\bf 28B}, 493 (1969). 

%
\bibitem{SeeSaw} T. Yanagida, in {\it Proceedings of the Workshop on Unified  
Theories and Baryon Number in the Universe} edited by A. Sawada and A. Sugamoto 
(KEK Report No.79-18, Tsukuba, 1979), p.95; Prog. Theor. Phys. {\bf 64}, 1103 (1980);  
M. Gell-Mann, P. Ramond and R. Slansky, in {\it Supergravity} edited by P. van 
Nieuwenhuizen and D.Z. Freedmann (North-Holland, Amsterdam 1979), p.315; 
R.N. Mohapatra and G. Senjanovic, Phys. Rev. Lett. {\bf 44}, 912 (1980).

\bibitem{Zee} A. Zee, Phys. Lett. {\bf 93B}, 389 (1980); {\bf 161B}, 141 (1985); 
L. Wolfenstein, Nucl. Phys. {\bf B175}, 93 (1980).

\bibitem{Babu} A. Zee, Nucl. Phys. {\bf 264B}, 99 (1986); 
K. S. Babu, Phys.  Lett. B {\bf 203}, 132 (1988); D. Chang, W-Y.Keung and P.B. Pal, 
Phys. Rev. Lett. {\bf 61}, 2420 (1988); 
J.T. Peltoniemi, A. Yu. Smirnov and J.W.F. Valle, Phys. Lett. B {\bf 286}, 321 (1992). 
For a model with $\nu_R$, see, for example,
D. Choudhury, R. Gandhi, J.A. Gracey and B. Mukhopadhyaya, Phys. Rev. D {\bf 50}, 3468 (1994). 

%
\bibitem{Radiative} For example, S. T. Petcov, Phys. Lett. {\bf 115B}, 401 (1982);    
K. S. Babu and V. S. Mathur, Phys.  Lett. B {\bf 196}, 218 (1987);   
J. Liu, Phys.  Lett. B {\bf 216}, 367 (1989); 
D. Chang and W.-Y. Keung, Phys. Rev. D {\bf 39}, 1386 (1989); 
W. Grimus and H. Neufeld, Phys.  Lett. B {\bf 237}, 521 (1990); 
B. K. Pal, Phys. Rev. D {\bf 44}, 2261 (1991); 
W. Grimus and G. Nardulli, Phys.  Lett. B {\bf 271}, 161 (1991); 
A. Yu. Smirnov and Z. Tao, Nucl. Phys. B {\bf 426}, 415 (1994).

%
\bibitem{ZeeType} 
    A. Yu. Smirnov and M. Tanimoto, Phys. Rev. D \textbf{55}, 1665 (1997);
    N. Gaur, A. Ghosal, E. Ma and P. Roy, Phys. Rev. D \textbf{58}, 071301 (1998);
    C. Jarlskog, M. Matsuda, S. Skadhauge and M. Tanimoto, Phys. Lett. B \textbf{449}, 240 (1999);
    Y. Okamoto and M. Yasu\`{e}, Prog. Theor. Phys. \textbf{101}, 1119 (1999);
    P. H. Frampton and S. L. Glashow, Phys. Lett. B \textbf{461}, 95 (1999);
    G. C. McLaughlin and J. N. Ng, Phys. Lett. B \textbf{455}, 224 (1999);
    A. S. Joshipura and S. D. Rindani, Phys. Lett. B \textbf{464}, 239 (1999);
    J. E. Kim and J. S. Lee, hep-ph/9907452 (July, 1999);
    N. Haba, M. Matsuda and M. Tanimoto, Phys. Lett. B \textbf{478}, 351 (2000);
    C-K. Chua, X-G. He and W-Y. P. Hwang, Phys. Lett. B \textbf{479}, 224 (2000);
    D. Chang and A. Zee, Phys. Rev. D \textbf{61}, 071303 (2000);
    K. Cheung and O. C. W. Kong, Phys. Rev. D \textbf{61}, 113012 (2000);

%
\bibitem{BiMaxmal} 
    D. V. Ahluwalia, Mod. Phys. Lett. A \textbf{13}, 2249 (1998);
    V. Barger, P. Pakvasa, T. J. Weiler and K. Whisnant, Phys. Lett. B \textbf{437}, 107 (1998);
    A. Baltz, A. S. Goldhaber and M. Goldhaber, Phys. Rev. Lett. \textbf{81}, 5730 (1998);
    M. Jezabek and Y. Sumino, Phys. Lett. B \textbf{440}, 327 (1998);
    R. N. Mohapatra and S. Nussinov, Phys. Lett. B \textbf{441}, 299 (1998);
    Y. Nomura and T. Yanagida, Phys. Rev. D \textbf{59}, 017303 (1999);
    Q. Shafi and Z. Tavartkiladze, Phys. Lett. B \textbf{451}, 129 (1999);
    \textit{ibid}. \textbf{482}, 145 (2000);
    I. Starcu and D.V.Ahluwalia, Phys. Lett. B \textbf{460}, 431 (1999);
    H. Georgi and S. L. Glashow, Phys. Rev. D \textbf{61}, 097301 (2000);
    R. N. Mohapatra, A. P\'{e}rez-Lorenzana and C. A. de S. Pires, Phys. Lett. B \textbf{474}, 355 (2000).

%
\bibitem{NearlyBiMaximal} H. Fritzsch and Z. Xing, Phys. Lett. B {\bf 372}, 265 (1996); 
M. Fukugida, M. Tanimoto and T. Yanagida, Phys. Rev. D {\bf 57}, 4429 (1998); 
M. Tanimoto, Phys. Rev. D. {\bf 59}, 017304 (1999).

%
\bibitem{LeLmuLtau} R. Barbieri, L.J. Hall, D. Smith, A. Strumia and N. Weiner, 
JHEP {\bf 12}, 017 (1998).  See also, 
S. T. Petcov, Phys. Lett. \textbf{110B}, 245 (1982);
C.N. Leung and S. T. Petcov, Phys. Lett. \textbf{125B}, 461 (1983);
A. Zee, in Ref.\cite{Babu}. 
%
\bibitem{SU3U1} 
F. Pisano and V. Pleitez, Phys. Rev. D {\bf 46}, 410 (1992); 
P.H. Frampton, Phys. Rev. Lett. {\bf 69}, 2889 (1992); 
D. Ng, Phys. Rev. D {\bf 49}, 4805 (1994). 
See also M. Singer, J.W.F Valle and J. Schechter, Phys. Rev. D {\bf 22}, 738 (1980).

%
\bibitem{OkamotoYasue} Y. Okamoto and M. Yasu${\grave {\rm e}}$, Phys. Lett. B {\bf 466}, 267 (1999).

%
\bibitem{EarlyRadiativeNu} R. Barbieri and R.N. Mohapatra, Phys. Lett. B {\bf 218},  225(1989); 
J. Liu, Phys. Lett. B {\bf 225},  148(1989).

%
\bibitem{RadiativeNu} V. Pleitez and M.D. Tonasse, Phys. Rev. D {\bf 48}, 5274 (1993); 
F. Pisano, V. Pleitez and M.D. Tonasse, hep-ph/9310230 v2 (Feb., 1994); 
P.H. Frampton, P.I. Krastev and J.T. Liu, Mod. Phys. Lett. {\bf A9}, 761 (1994); 
M.B. Tully and G.C. Joshi, hep-ph/9810282 (Oct., 1998). See also J.W.F Valle and M. Singer, 
Phys. Rev. D {\bf 28}, 540 (1983).

%
\bibitem{TreeNu} 
For mechanisms of well-suppressed tree-lever neutrino masses, 
see M.B. Tully and G.C. Joshi, hep-ph/0011172 (Nov., 2000); 
J.C. Montero, C.A. Pires and V. Pleitez, hep-ph/0011296 (Nov., 2000). 

%
\bibitem{BabuType} 
For two-loop radiative models with the $L^\prime$ conservation, see 
    L. Lavoura, Phys. Rev. D {\bf 62}, 093011 (2000);
    T. Kitabayashi and M. Yasu\`{e}, Phys. Lett. B {\bf 490}, 236 (2000).

%
\bibitem{OtherVariant} For a model with ($\phi^0$, $\phi^-$, $h^+$), see T. Kitabayashi and M. Yasu${\grave {\rm e}}$, hep-ph/0010087 (Oct., 2000).

%
\bibitem{HeavyLepton} 
M. Singer, J.W.F Valle and J. Schechter, in Ref.\cite{SU3U1}; 
F.-z. Chen, Phys. Lett. B {\bf 442}, 223 (1998); 
V. Pleitez and M.D. Tonasse, Phys. Lett. B {\bf 430}, 174 (1998).

%
\bibitem{OneTwoLoop} A.S. Joshipura and S.D. Rindani, in Ref.\cite{ZeeType}; 
D. Chang and A. Zee, in Ref.\cite{ZeeType}.  See also, 
    J. T. Peltoniemi, A. Yu. Smirnov and J. W. F. Valle, in Ref. \cite{Babu};
    J. T. Peltoniemu, D. Tommasini and J. W. F. Valle, Phys. Lett. B {\bf 298}, 383 (1993); 
    J. T. Peltoniemu and J. W. F. Valle, Nucl. Phys. B {\bf 406}, 409 (1993). 


%
\bibitem{Simplest} The simple extension would be to use $\ell_R\ell_Rk^{++}$ (as in \cite{Babu}) and 
$\rho^{\prime +\dagger}\rho^{\prime +\dagger}k^{++}$. 
However, it is impossible to form an 
$SU(3)\times U(1)_N$-singlet that contains $\rho^{\prime +}\rho^{\prime +}$. 

%
\bibitem{Duplicate} R. Barbieri and R.N. Mohapatra, in Ref.\cite{EarlyRadiativeNu}.

%
\bibitem{FCNC} 
S.L. Glashow and S. Weinberg, Phys. Rev. D {\bf 15}, 1985 (1977); 
H. Georgi and A. Pais, Phys. Rev. D {\bf 19}, 2746 (1979).

%
\bibitem{FCNCSU3} 
J.C. Montero, F. Pisano and V. Pleitez, Phys. Rev. D {\bf 47}, 2918 (1993); 
M. $\ddot{{\rm O}}$zer, Phys. Rev. D {\bf 54}, 4561 (1996); 

%
\bibitem{ZeeModelNU}  C. Jarlskog, M. Matsuda, S. Skadhauge and M. Tanimoto, in Ref.\cite{ZeeType}. 
See also, A.Yu. Smirnov and M. Tanimoto, in Ref.\cite{ZeeType}.

%
\bibitem{RecentAtm} For recent analysis, see N. Fornengo, M.C. Gonzalez-Garcia and J.W.F. Valle, 
JHEP {\bf 7}, 006 (2000); Nucl. Phys. B {\bf 580}, 58 (2000). 

%
\bibitem{VOComment}
The announcement that the VO solution seems to be disfavored at the 95$\%$ confidence level has been made by the SuperKamiokande collaboration \cite{RecentSK}.  However, it is stressed that this statement is not conclusive\cite{Takeuchi} and that theorists keep watching the VO solution to solar neutrino problem. See also C.E.C Lima, H.M. Portella and L.C.S. de Oliveira, hep-ph/0010038 v2 (Oct., 2000); R. Barbieri and A. Strumia, hep-ph/0011307 v2 (Nov., 2000).

%
\bibitem{Takeuchi}
    Y. Takeuchi, a talk at {\it Post Summer Institute 2000 on Neutrino Physics}, August 21-24, 2000, Fuji-Yoshida, 
Japan.

%
\bibitem{Appendix} T. Kitabayashi and M. Yasu${\grave {\rm e}}$, in Ref.\cite{BabuType}.
\end{references}
\end{document}